\documentclass{SCIS2021}

\usepackage{balance}
\usepackage{textcomp}
\usepackage{booktabs}
\usepackage{enumitem}
\usepackage{multirow}
\usepackage{subfig}
\usepackage{listings}
\usepackage{color}
\usepackage{pifont}
\usepackage{tcolorbox}
\usepackage{threeparttable}

\usepackage{dingbat}
\usepackage{bbding}

\newcommand{\xmark}{\ding{55}}%

\begin{document}

\ArticleType{RESEARCH PAPER}
\Year{2020}
\Month{December}
\Vol{}
\No{}
\DOI{}
\ArtNo{}
\ReceiveDate{}
\ReviseDate{}
\AcceptDate{}
\OnlineDate{}


\title{How does Working from Home Affect Developer Productivity? \\ \Large{-- A Case Study of Baidu During the COVID-19 Pandemic}}

\author[1]{Lingfeng Bao}{}
\author[2]{Tao Li}{}
\author[3]{Xin Xia}{{Xin.Xia@monash.edu}}
\author[2]{Kaiyu Zhu}{}
\author[2]{Hui Li}{}
\author[1]{Xiaohu Yang}{}

\AuthorMark{Lingfeng Bao}

\AuthorCitation{Bao et al. }

\address[1]{Zhejiang University, China, Hangzhou, China}
\address[2]{Baidu Inc., Beijing, China}
\address[3]{Monash University, Melbourne, Australia}

\abstract{Nowadays, working from home (WFH) has become a popular work arrangement due to its many potential benefits for both companies and employees (e.g., increasing job satisfaction and retention of employees). Many previous studies have investigated the impact of working from home on the productivity of employees. However, most of these studies usually use a qualitative analysis method such as surveys and interviews, and the studied participants do not work from home for a long continuing time. Due to the outbreak of coronavirus disease 2019 (COVID-19), a large number of companies asked their employees to work from home, which provides us an opportunity to investigate whether working from home affects their productivity.

In this study, to investigate the difference of developer productivity between working from home and working onsite, we conduct a quantitative analysis based on a dataset of developers' daily activities from Baidu Inc, one of the largest IT companies in China. In total, we collected approximately four thousand records of 139 developers' activities of 138 working days. Out of these records, 1,103 records are submitted when developers work from home due to the COVID-19 pandemic. We find that WFH has both positive and negative impacts on developer productivity in terms of different metrics, e.g., the number of builds/commits/code reviews. We also notice that working from home has different impacts on projects with different characteristics including programming language, project type/age/size. For example, working from home has a negative impact on developer productivity for large projects. Additionally, we find that productivity varies for different developers. Based on these findings, we get some feedback from developers of Baidu and understand some reasons why WFH has different impacts on developer productivity. We also conclude several implications for both companies and developers.
}


\maketitle

\section{Introduction}\label{sec:intro}
Working from home (WFH) is a work arrangement in which employees do not need not work at a central place (e.g., office building, warehouse, or store).
Working from home has various names, such as remote work, teleworking, or telecommuting. These terms are used differently and interchangeably from study to study~\cite{shin2000telework, spark2017accessibility, hill2003does}.
Nowadays, since working from home is facilitated by many tools such as virtual private network, cloud computing, and online meeting software, 
more and more companies allow their employees to work from home.  
A survey in 2018 from OWL labs shows that 52\% of employees work from home at least once a week and 56\% of companies allow remote work\footnote{https://www.owllabs.com/state-of-remote-work/2018}. 
Working from home can offer some benefits to both companies and employees, for instance, when employees can work from home, they feel more trusted and are better able to balance work and life responsibilities, which can increase employee retention and make them happier and more productive.

Working from home is also adopted by many IT companies, for instance, a recent report mentioned that Twitter announced staff can continue working from home permanently\footnote{https://www.bbc.com/news/technology-52628119}.
Developers can perform their daily tasks (e.g., writing code, debugging, build projects, and code review) as usual by remotely accessing resources of companies when working from home. 
Working from home might have different impacts on productivity, which is a big concern of software developer organizations~\cite{meyer2017work}. 
Understanding the difference of developer productivity when working from home and the reasons behind it can help improve the management of companies and projects, increase the job satisfaction of developers, and make developers more productive.

The survey of OWL labs reported that employees who work remotely at least once a month are 24\% more likely to feel productive in their roles than those who do not or cannot work remotely.
On the contrary, Working from home might have a negative impact on productivity. For example, it would decrease the efficiency of developer communication, which plays an important role in software development~\cite{wolf2009predicting}. 
Many studies have investigated the impacts of working from home on productivity~\cite{neufeld2004predicting, neufeld2005individual, baker2007satisfaction, laihonen2012measuring, campbell2015flexible}. 
However, most of these studies use a qualitative approach (e.g., survey or interview) based on the feedback from general workers (not only developers). Additionally, the studied participants usually do not work from home for a long continuing time. 
In this study, we aim to investigate the impacts of developer productivity when working from home for a long time in a quantitative way.

Due to the outbreak of coronavirus disease 2019 (COVID-19)\footnote{https://www.who.int/health-topics/coronavirus}, which is an infectious disease caused by a newly discovered coronavirus, a large number of IT companies ask their employees to work from home,
which provides us an opportunity to investigate how their productivity is affected when working from home for a long continuing time. 

In this study, we collect the data from Baidu, Inc., China, which contains the development activities from 107 developers in 70 working days. There is a part of records in this dataset on which developers work from home due to the COVID-19 pandemic.
We compare developer productivity when working from home with working onsite in terms of multiple aggregated values such as \emph{mean}, \emph{median} of several metrics (e.g., the number of builds, commits, and code reviews.). 
We summarize our findings and contributions as follows:
\begin{itemize}[leftmargin=*]
    \item To the best of our knowledge, we are the first to investigate the impacts of working from home on developer productivity based on developers' daily activities. We find that working from home has both positive and negative impacts on developer productivity in terms of different metrics, such as the number of builds/commits/code reviews. 
    \item We investigate the impacts of working from home on projects with different characteristics including program language and project type/age/size and find that working from home has different impacts on different kinds of projects. For example, working from home has a negative impact on developer productivity for large projects. We also find that productivity varies for different developers. 
    \item We conclude the reasons why developers have different productivity when working from home and provide implications based on our findings and the feedback from Baidu. 
\end{itemize}


\vspace{0.1cm}  \noindent {\bf Paper Structure:}
The remainder of the paper is structured as follows. Section~\ref{sec:setup} describes the dataset and research questions in this study. Section~\ref{sec:results} presents the results of the analysis for the six research questions. Section~\ref{sec:diss} discusses implications and threats to validity. Section~\ref{sec:related} briefly reviews related
work. Section~\ref{sec:conclusion} concludes the paper and discusses future directions.

\section{Case Study Setup}\label{sec:setup}
In this section, we first present the dataset from Baidu. Then, we describe the research questions and their corresponding motivations. 

\subsection{Dataset}\label{sec:dataset}


\begin{table}[]
    \centering
    \caption{The overview of the dataset}\label{tbl:wfh}
\begin{tabular}{@{}llrr@{}}
    \toprule
                               & Date Range   & \#Working Days           & \#Records \\ \midrule
    {\tt DATA\_ONSITE} & 2019/12/23 - 2020/02/02 & 42 & 1,325      \\
    {\tt DATA\_WFH}         & 2020/02/03 - 2020/03/01 & 28 & 1,103      \\ \midrule
    Total                      &             &   70         & 2,428      \\ \bottomrule
\end{tabular}
\end{table}


We collected a dataset of developers' daily activities from Baidu, Inc., which is the world's largest Chinese language Internet search provider\footnote{https://www.baidu.com/. As the world's largest Chinese language Internet search provider, Baidu responds to a huge amount of search queries from more than 100 countries and regions every day, serving as the most important way for netizens to access Chinese language information. With its mission to ``make the complicated world simpler through technology'', Baidu promotes constant technological innovation, and is committed to being a world-leading technology company that understands its users and helps them grow.}, the largest knowledge and information centered Internet platform company in China, and a world-leading artificial intelligence (AI) company. 

The dataset we get from Baidu contains 107 developers' daily activities from eight projects in 70 working days.
Table~\ref{tbl:wfh} presents the overview of the dataset. As shown in this table, the time of working from home is from 2020/02/03 to 2020/03/01 because Baidu asked all its employees to work from home after the outbreak of the COVID-19 pandemic in China\footnote{Wuhan is the first city in China to be lockdown since 2020/01/23, and the whole country started to lockdown from 2020/01/27, which is during the Spring Festival (2020/01/25) -- the most important holiday in China. After the holiday, Baidu asked its employees to work from home.}. We refer this part of records as to {\tt DATA\_WFH}. 
On the other hand, we also got a list of records during which developers work onsite, i.e., from 2019/12/23 to 2020/02/02 (referred to as {\tt DATA\_ONSITE}). 
The development activities in the dataset {\tt DATA\_WFH} and {\tt DATA\_ONSITE} are similar since they happened very closely. 
This dataset has a total of 2,428 records. Among these records, there are 1,325 and 1,103 records that are belong to {\tt DATA\_ONSITE} and {\tt DATA\_WFH}, respectively.

\begin{table}[]
    \centering
    \caption{The metrics of each record of developers' daily activities provided by Baidu.}\label{tbl:metrics}
    \begin{tabular}{@{}ll@{}}
    \toprule
    Feature              & Description                                         \\ \midrule
    date\_partition      & The date of a record                                       \\
    username\_e          & The encrypted user name of a developer              \\
    repo\_name\_e        & The encrypted repository name                       \\
    commit\_count        & The number of commits submitted by a developer      \\
    line\_inserted       & The number of lines of code inserted by a developer \\
    line\_deleted        & The number of lines of code deleted by a developer  \\
    review\_count        & The number of code reviews performed by a developer  \\
    job\_status\_build   & The status of build performed by a developer       \\
    build\_count         & \begin{tabular}[c]{@{}l@{}}The number of builds performed by a developer;\\ a build refers to the process of continuous integration, \\ including compilation, test, deploy, etc.\end{tabular} \\
    job\_status\_release & The status of release performed by a developer    \\
    release\_count       & The number of releases performed by a developer     \\
    compile\_count       & The number of compilations performed by a developer     \\ \bottomrule
    \end{tabular}
\end{table}

Each record in the dataset has several metrics that represent the activities of a developer in one day.  
Table~\ref{tbl:metrics} presents the fields of a record. Each record has a date (\emph{data\_partition}) on which a developer's activities are reported. The developers' names and their project in the record are encrypted into unique IDs due to the security and privacy policy of Baidu so that we can still track records over time. 
Each record has the following numeric metrics: \emph{commit\_count}, \emph{line\_inserted}, \emph{line\_deleted}, \emph{review\_count}, \emph{build\_count}, \emph{release\_count}, \emph{compile\_count}.
Although these numeric metrics are dependent on various factors such as developer experience, programming languages and styles~\cite{kamei2012large}, many previous studies have used similar quantitative metrics such as lines of code to measure developer productivity~\cite{walston1977method, devanbu1996analytical, nguyen2011analysis}. Therefore, we believe these metrics can potentially indicate developer productivity.

Due to the security policy of Baidu, the numeric metrics are standardized by the following formula:
$ z = \frac{X-\mu}{\sigma}$,
where $z$ is the standardized value, $X$ is the real value of a metric in a record, $\mu$ is the mean of a metric in the dataset, and $\sigma$ is the standard deviation of a metric in the dataset. Thus, the standardized values do not affect the distribution of a metric and the findings in this study since the findings are based on the comparison between the values of metrics when developers work from home and work onsite. 
Moreover, the standardized values can be positive or negative.
Additionally, there are two other fields, i.e., \emph{job\_status\_build} and \emph{job\_status\_release}, which represent the status of builds/releases performed by a developer. There are four possible values for the status of builds/releases, including success, failed, canceled, or NULL. Comparing to the other three statuses, a successful build or release means that a developer is more productive on that day. 


\begin{table}[]
\centering
\caption{The project information.}\label{tbl:projects}
\begin{tabular}{@{}llllr@{}}
\toprule
   & Created Year & Project Type    & Language   & \#Developer \\ \midrule
P1 & 2017        & APP               & C++        & 13          \\
P2 & 2017        & APP              & Java & 13          \\
P3 & 2018        & SERVER             & Java & 4           \\
P4 & 2017        & SERVER         & Java & 25          \\
P5 & 2018        & SERVER             & C++        & 53          \\
P6 & 2018        & SDK                & C++        & 10          \\
P7 & 2018        & SERVER             & Java & 7           \\
P8 & 2017        & SERVER         & C++        & 14          \\ \bottomrule
\end{tabular}
\end{table}

Table~\ref{tbl:projects} presents the information about the eight projects in the dataset. These projects are created in two different years, i.e., 2017 and 2018. There are three types of projects, i.e., APP (application software), SERVER (server software such as web services, API libraries), and SDK (software development kits).
Among these projects, four projects are written in C++ while the other four projects are written in Java. 
Additionally, we count the number of developers who have records of development activities in the dataset for each project. 
We also report that these projects have different numbers of developers, for example, project P5 has the most number of developers (i.e., 53) and P3 has the least number of developers (only 4 developers).



\subsection{Research Questions}
In this section, we present the six research questions we address in our study:

\vspace{0.1cm}  \noindent {\bf RQ1. Are there any significant differences between the productivity of developers working from home and working onsite?}

 \noindent {\bf Motivation:}
In this RQ, we want to investigate whether working from home can affect developers' productivity comparing to working onsite. Given the dataset from Baidu, we measure the overall productivity of all developers by combining their activities together, then compare the overall productivity when working from home and working onsite.


\vspace{0.1cm}  \noindent {\bf RQ2. Do different programming languages affect developer productivity when working from home?}

 \noindent {\bf Motivation:}
Previous studies have shown that programming languages have an important impact on developers' activities, such as programming comprehension~\cite{xia2017measuring}, and being a long-time contributor of open source projects~\cite{von2003community, bao2019large}. 
Working from home might have different impacts on developers using different programming languages. 
For example, since C++ projects in Baidu are usually larger and require more computing resources than Java projects, developers often need to build and debug these C++ projects on a powerful machine remotely.
Meanwhile, for most Java projects, developers can write code and debug in their own computers at home.
Thus, in this RQ, we want to investigate whether developers using different programming languages have different productivities when working from home.

\vspace{0.1cm}  \noindent {\bf RQ3. Do different project types affect developer productivity when working from home?}

  \noindent {\bf Motivation:}
As shown in Table~\ref{tbl:projects}, the eight projects in the dataset have three different types, i.e., APP, SERVER, and SDK. 
The projects with different project types could have different project management methods and styles of schedules, different development and communication tools, which might have a potential impact on developers' productivity~\cite{melo2011agile, meyer2014software}. 
Working from home changes the way of project management and development, which have different impacts on developer productivity for projects with different types. 
For example, many APP projects develop mobile apps, which usually rely on some specific framework and have predefined code styles and specifications. While the software developed by SERVER projects are usually applied in much more complicated scenarios and depend on many different frameworks and programming languages.
Thus, it might be more different for developers when working from home to build, test, and debug a SERVER project than a APP project.
Thus, we want to investigate whether different project types have an impact on developer productivity when working from home.

\vspace{0.1cm}  \noindent {\bf RQ4. Do different project ages affect developer productivity who working from home?}

  \noindent {\bf Motivation:}
Different project ages might affect developers' activities. For example, our previous study found that developers in older projects spend more time on program comprehension activities than those of projects in the beginning stage~\cite{xia2017measuring}. 
Working from home might expand such effects caused by project age, for example, for an older project, developers need to read the source code and documents more frequently since such a project usually has more maintenance tasks; but they cannot access these resources and communicate with colleagues easily when working from home, which might lower their productivity. 
Thus, we want to investigate whether different project phases have an impact on the productivity of developers when working from home in this RQ.

\vspace{0.1cm}  \noindent {\bf RQ5. Do different project sizes affect developer productivity when working from home?}

 \noindent {\bf Motivation:}
Different project sizes (measured by the number of developers of a project in this study) might have an impact on developers' productivity~\cite{paiva2010factors}. 
Zhou et al. find that size as a factor has been always considered as a confounding effect in different approaches in software engineering~\cite{zhou2009examining}.
Due to different project sizes, projects might have different ways of project management and software development, which would be affected by working from home differently. 
For example, it might be more difficult to communicate with each other in a large project when working from home, which might decrease the productivity of the project.

\vspace{0.1cm}  \noindent {\bf RQ6. Do individual developers have a different productivity when working from home?}

 \noindent {\bf Motivation:}
Comparing with working onsite, developers might have different productivity when working from home due to some personal factors, e.g., experience, personality, habit, and skills. 
For example, developers are much easier to be interrupted by some other stuff when working from home.
In RQ1 we use the aggregated activity data of developers while in this RQ we user the daily activities of individual developers to investigate their productivity. 
Identifying developers who have different performance when working from home and the potential reasons behind it can help project leaders manage the projects. 
Thus, in this RQ we want to investigate whether individual developers have different productivity when working from home.

\begin{figure}[t!]
    \centering
    \includegraphics[width=0.8\textwidth]{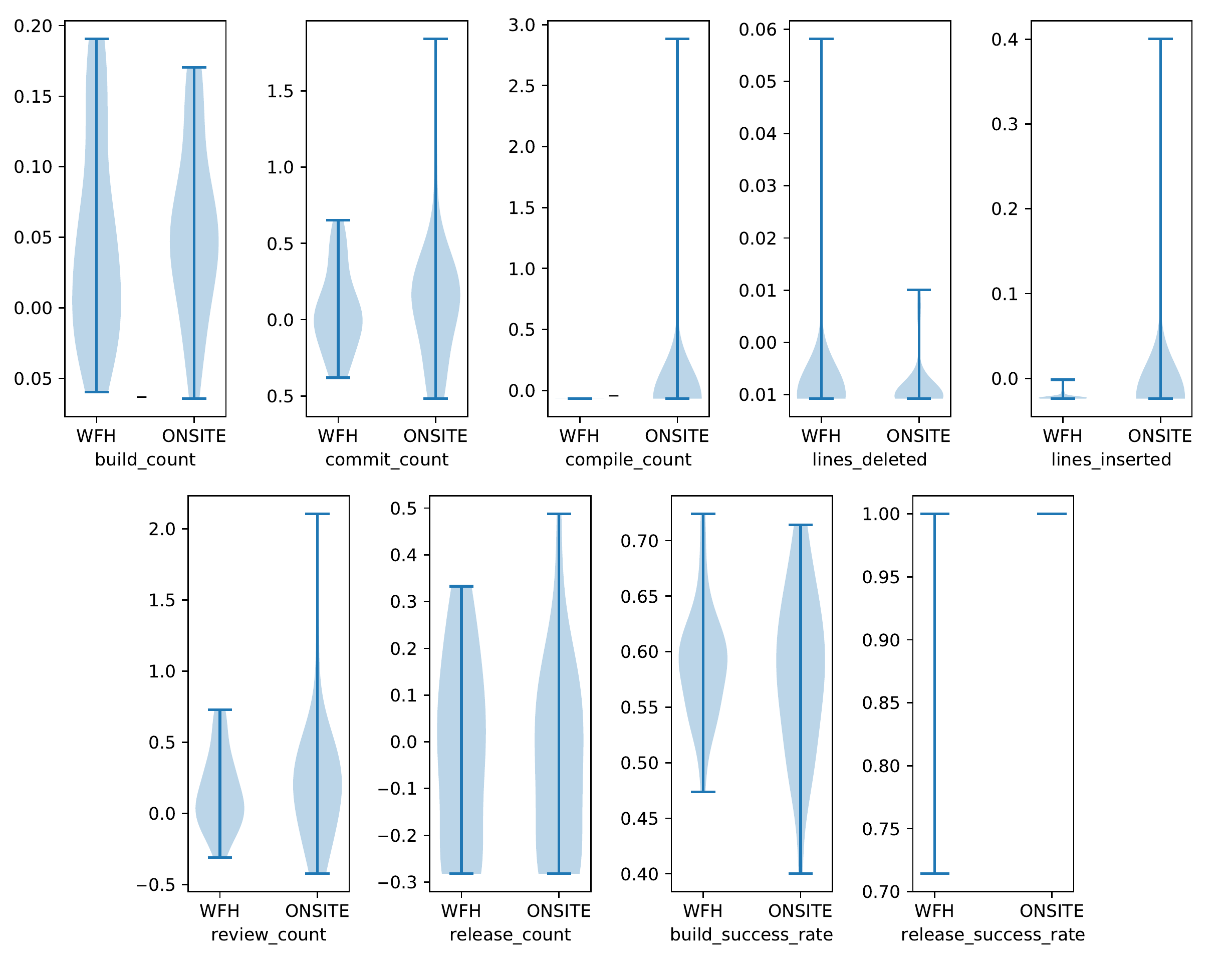}
    \caption{Violin plots for {\tt DATA\_WFH} and {\tt DATA\_ONSITE} in terms of the mean of numeric metrics grouped by day and build/release successful rate in one day.}
    \label{fig:violin}
\end{figure}

\section{Case Study Results}\label{sec:results}

In this section, we present the results of these six research questions one by one.

\subsection{RQ1. The Overall Developer Productivity when Working From Home}\label{sec:rq1}

\vspace{0.1cm}  \noindent {\bf Methodology:}
To compare developer productivity when working from home with that when working onsite, we first group the records in the dataset by day, and compute several aggregate values including \emph{mean}, \emph{median}, \emph{sum}, \emph{max}, and \emph{min} for each numeric metric, as shown in Table~\ref{tbl:projects}. Thus, we can know a more accurate distribution of developers' productivity in terms of each metric. For example, although \emph{mean} of the submitted commits (\emph{commit\_count}) in one day indicates the average workload of developers in one day, but sometimes a few experienced/core developers usually contribute more commits than junior/periphery developers, the mean of commits might still be very high. For all the metrics except for \emph{sum}, we calculate these aggregate values of each metric of all developers for each day. 
For \emph{sum}, we only consider the developers whose records are both in {\tt DATA\_WFH} and {\tt DATA\_ONSITE}, i.e., the sum of {\tt DATA\_WFH} and {\tt DATA\_ONSITE} includes the same number of developers.
For the two non-numeric feature \emph{job\_status\_build} and \emph{job\_status\_release}, we compute the success rate of build/release in one day, that is, the ratio of the number of times that a build/release is successful.

\begin{table*}[]
    \centering
    \caption{P-value and Cliff's delta ($\delta$) for {\tt \textbf{DATA\_WFH}} compared with {\tt \textbf{DATA\_ONSITE}} in terms of different aggregate values of each metric in one day.}\label{tbl:rq12}
    \resizebox{\textwidth}{!}{%
    \begin{tabular}{@{}lcccccccccc@{}}
    \toprule
    \multicolumn{1}{c}{\multirow{2}{*}{\textbf{Metric}}} & \multicolumn{2}{c}{\textbf{mean}} & \multicolumn{2}{c}{\textbf{median}} & \multicolumn{2}{c}{\textbf{sum}}  & \multicolumn{2}{c}{\textbf{max}}  & \multicolumn{2}{c}{\textbf{min}}    \\ \cmidrule(l){2-11} 
    \multicolumn{1}{c}{}                                  & \textbf{p-value} & \textbf{$\delta$} & \textbf{p-value}  & \textbf{$\delta$}  & \textbf{p-value} & \textbf{$\delta$} & \textbf{p-value} & \textbf{$\delta$} & \textbf{p-value} & \textbf{$\delta$} \\ \midrule
    build\_count    & 0.112 & 0.173  & \textbf{0.001} & \colorbox{blue!30}{-0.419} & 0.153 & 0.146  & \textbf{0.012} & \colorbox{blue!10}{0.320}  & \textbf{0.002} & \colorbox{blue!10}{-0.262} \\
    commit\_count   & 0.220 & 0.111  & \textbf{0.034} & \colorbox{blue!10}{0.245}  & 0.227 & 0.107  & \textbf{2.05E-04} & \colorbox{blue!60}{0.500}  & NA    & NA     \\
    compile\_count  & \textbf{0.036} & \colorbox{blue!10}{-0.225} & 0.216 & -0.024 & \textbf{0.027} & \colorbox{blue!10}{-0.273} & 0.216 & -0.024 & NA    & NA     \\
    lines\_deleted  & \textbf{0.005} & \colorbox{blue!30}{-0.371} & 0.139 & 0.152  & \textbf{0.006} & \colorbox{blue!30}{-0.359} & 0.378 & 0.045  & NA    & NA     \\
    lines\_inserted & \textbf{0.006} & \colorbox{blue!30}{-0.357}  & 0.175 & 0.132  & \textbf{0.007} & \colorbox{blue!30}{-0.350} & 0.172 & 0.135  & NA    & NA     \\
    release\_count  & 0.131 & 0.160 & NA    & NA     & 0.119 & 0.168  & \textbf{0.027} & \colorbox{blue!10}{0.259}  & NA    & NA     \\
    review\_count   & 0.144 & 0.151  & 0.075 & 0.182  & 0.161 & 0.141  & \textbf{0.015} & \colorbox{blue!10}{0.308}  & NA    & NA     \\ \bottomrule
   \end{tabular}
    }
\end{table*}

For each aggregation function of a metric, we have two groups, i.e., the days on which developers work from home and work onsite, respectively. The number of data points of a group is equal to the number of days of the corresponding group. 
Then, we apply the Wilcoxon rank-sum test~\cite{wilcoxon1945individual} to investigate whether the difference is statistically significant in terms of one kind of aggregate values of each metric. 
We also compute Cliff's delta~\cite{cliff2014ordinal}\footnote{Cliff defines a delta of less than 0.147, between 0.147 to 0.33, between 0.33 and 0.474, and above 0.474 as negligible, small, medium, and large effect size, respectively.} to quantify the amount of difference between the two groups. 
Consequently, we can compare developer productivity when working from home with working onsite in terms of different metrics.

\vspace{0.1cm}  \noindent {\bf Results:}
Since we only have standardized values for these metrics in the dataset, we use violin plots to show the distributions of these metrics when working from home and working onsite, as shown in Figure~\ref{fig:violin}. 
Notice that the values of these metrics have been standardized (see Section~\ref{sec:dataset}), their values can be negative.
From this figure, these metrics have less wide distributions when developers work from home than those when developers work onsite except the number of lines deleted every day. For example, the range of the mean of \emph{commit\_count} by day when working onsite is approximately from -0.4 to 0.6, while the range of the mean of \emph{build\_count} by day when working from home is approximately from -0.5 to 1.8. This indicates that developers might have more stable productivity when working from home than when they work onsite. 

Table~\ref{tbl:rq12} presents p-values and Cliff's deltas ($\delta$) for {\tt DATA\_WFH} compared with {\tt DATA\_ONSITE} in terms of different aggregate values of each metric in one day. 
There are some cases whose values are ``NA'' in these tables, which is caused by the two compared lists are completely the same.
We find that there are some cases in which developer productivity when working from home is significantly different from that when working onsite. 
To ease inspection, we highlight those Cliff's delta values when the p-values are significant in different background colors for different levels of the effect size.
For example, the sum of \emph{compile\_count}, \emph{line\_deleted} and \emph{line\_inserted} when working from home is significantly less than those when working onsite, and all the effect sizes are not at the negligible level. This might indicate Working from home has a negative effect on developer productivity in terms of the number of compilations, deleted lines, and inserted lines every day. 
We also compute p-values and Cliff's deltas for the success rate of build/release for {\tt DATA\_WFH} compared with {\tt DATA\_ONSITE}. 
We find that the success rate of build when working from home is not significantly different from when working onsite ($p-value=0.207$), while the success rate of release when working from home is significantly less from when working onsite ($p-value=0.002$ and $\delta=-0.842$). This might be because a project release require more collaboration and communication that are affected by working from home.  

Moreover, we find that working from home has different impacts on developer productivity in terms of different metrics. 
For example, in terms of \emph{build\_count}, its median values for developers when working from home is less than these when working onsite. On the contrary, the maximum value of \emph{build\_count} for developers when working from home is larger than these when working onsite.
This might indicate that a small number of developers perform more builds when working from home, while most of the developers perform fewer builds.
As shown in the results of RQ6 (see Section~\ref{sec:rq6}), this might be because some developers can be more productive when working from home. 
\begin{tcolorbox}
Overall, working from home has different impacts on developer productivity in terms of different metrics. 
\end{tcolorbox}

\begin{table}[]\vspace{-0.2cm}
    \centering
    \caption{The cases in which values when working from home is significantly different for C++ and Java projects}\label{tbl:language}
    \begin{tabular}{@{}llcc@{}}
    \toprule
    Language              & Metric          & Positive  & Negative               \\ \midrule
\multirow{7}{*}{C++}  & build\_count    & sum, max  & min                    \\
                      & commit\_count   & mean, max & ---                    \\
                      & compile\_count  & ---       & mean, sum              \\
                      & lines\_deleted  & ---       & sum                    \\
                      & lines\_inserted & ---       & sum                    \\
                      & release\_count  & mean      & ---                    \\
                      & review\_count   & mean, max & ---                    \\ \midrule
\multirow{7}{*}{Java} & build\_count    & ---       & mean, median, sum, min \\
                      & commit\_count   & ---       & sum                    \\
                      & compile\_count  & ---       & ---                    \\
                      & lines\_deleted  & ---       & ---                    \\
                      & lines\_inserted & ---       & ---                    \\
                      & release\_count  & ---       & ---                    \\
                      & review\_count   & ---       & mean                   \\ \bottomrule 
    \end{tabular}
    \end{table}

\subsection{RQ2. The Impact of Programming Language}\label{sec:rq2}

\vspace{0.1cm}  \noindent {\bf Methodology:}
For both the records in {\tt DATA\_WFH} and {\tt DATA\_ONSITE}, we divide them into two groups: those from the projects using C++ and Java.
For each group, we use the same method in the RQ1 then investigate whether the difference between the productivity of developers working from home and working onsite is statistically significant in terms of the aggregate values of each metric. 

\vspace{0.1cm}  \noindent {\bf Results:}
Table~\ref{tbl:language} presents the cases in which values when working from home are significantly different and the effect sizes are not negligible from these when working onsite for C++ and Java projects. 
The column `Positive'/`Negative' means that working from home has a positive/negative impact on developer productivity in terms of an aggregate value of a metric.
``---'' means there are no cases in which values when working from home are significantly different from these when working onsite for a metric. 

We find that for C++ projects, there are both some positive and negative cases. 
For example, the \emph{mean} of \emph{commit\_count}, \emph{release\_count}  and \emph{review\_count} are belong to positive cases while the \emph{sum} of \emph{compile\_count}, \emph{lines\_deleted} and \emph{lines\_inserted} are belong to negative cases. 
On the other hand, all the cases of {\tt Java} projects belong to negative cases. 
This indicates that working from home has more negative impacts on Java projects than C++ projects. 



\begin{tcolorbox}
    Working from home has both positive and negative impacts on developer productivity for C++ projects in terms of different metrics but has a negative impact on developer productivity for Java projects.
\end{tcolorbox}

\begin{table}[]
    \centering
    \caption{The cases in which values when working from home is significantly different for {\tt \textbf{APP}}, {\tt \textbf{SDK}} and {\tt \textbf{Server}} projects}\label{tbl:projecttype1}
    \begin{tabular}{@{}llcc@{}}
        \toprule
        Type                    & Metric          & Positive               & Negative    \\ \midrule
        \multirow{7}{*}{APP}    & build\_count    & max                    & mean, min   \\
                                & commit\_count   & mean, median, sum, max & ---         \\
                                & compile\_count  & ---                    & mean        \\
                                & lines\_deleted  & ---                    & sum         \\
                                & lines\_inserted & ---                    & ---         \\
                                & release\_count  & mean                   & ---         \\
                                & review\_count   & mean, median, sum, max & ---         \\ \midrule
        \multirow{7}{*}{SDK}    & build\_count    & mean, max              & min         \\
                                & commit\_count   & ---                    & min         \\
                                & compile\_count  & ---                    & ---         \\
                                & lines\_deleted  & ---                    & min         \\
                                & lines\_inserted & ---                    & min         \\
                                & release\_count  & ---                    & ---         \\
                                & review\_count   & ---                    & min         \\ \midrule
        \multirow{7}{*}{SERVER} & build\_count    & ---                    & median, min \\
                                & commit\_count   & ---                    & ---         \\
                                & compile\_count  & ---                    & mean, sum   \\
                                & lines\_deleted  & ---                    & sum         \\
                                & lines\_inserted & ---                    & sum         \\
                                & release\_count  & ---                    & mean        \\
                                & review\_count   & ---                    & ---         \\ \bottomrule
    \end{tabular}
    \end{table}



\subsection{RQ3. The Impact of Project Type}

\vspace{0.1cm}  \noindent {\bf Methodology:}
We split the records in the dataset into three parts based on the project type, i.e., APP, SERVER, and SDK. 
For each part, we use the same approach as RQ1 and RQ2 to investigate the difference between the productivity of developers when working from home and working on site for projects with different types.

\vspace{0.1cm}  \noindent {\bf Results:}
Table~\ref{tbl:projecttype1} presents the cases in which values when working from home are significantly different from those when working onsite and and the effect sizes is not negligible for APP, SERVER, and SDK projects. 
For APP projects, there are some positive and negative cases, which indicate that working from home might have both positive and negative impacts on developer productivity. For example, in terms of \emph{mean}, \emph{median}, \emph{sum} and \emph{max} of \emph{commit\_count} and \emph{review\_count}, their values for developers when working from home are significantly larger than these when working onsite; on the contrary, in terms of \emph{mean} of \emph{build\_count} and \emph{compile\_count}, their values for developers when working from home are significantly less than these when working onsite.

For SDK and SERVER projects, most of the cases are negative, which indicates that working from home has a negative impact on developer productivity. 
This might be because SDK and SERVER projects usually have more components than APP projects, which requires more collaboration and communication with the other developers. According to the feedbacks from Baidu (see Section~\ref{sec:feedbacks}), working from home has a negative impact on collaboration and communication, which decreases developer productivity.



\begin{tcolorbox}
    Working from home has both positive and negative impacts on developer productivity for APP projects and decreases developer productivity for SDK and SERVER projects.
\end{tcolorbox}



\begin{table}[]\vspace{-0.3cm}
    \centering
    \caption{The cases in which values when working from home is significantly different for projects created in 2017 and 2018}\label{tbl:age}
    \begin{tabular}{@{}llcc@{}}
    \toprule
    Year                  & Metric          & Positive               & Negative    \\ \midrule
    \multirow{7}{*}{2017} & build\_count    & max                    & median, min \\
                          & commit\_count   & mean, sum, max         & ---         \\
                          & compile\_count  & ---                    & mean        \\
                          & lines\_deleted  & median, max            & ---         \\
                          & lines\_inserted & median, max            & sum         \\
                          & release\_count  & mean                   & ---         \\
                          & review\_count   & mean, median, sum, max & ---         \\ \midrule
    \multirow{7}{*}{2018} & build\_count    & ---                    & median, min \\
                          & commit\_count   & ---                    & ---         \\
                          & compile\_count  & ---                    & mean, sum   \\
                          & lines\_deleted  & ---                    & mean, sum   \\
                          & lines\_inserted & ---                    & sum         \\
                          & release\_count  & ---                    & ---         \\
                          & review\_count   & ---                    & sum         \\ \bottomrule
    \end{tabular}\vspace{-0.3cm}
\end{table}

\subsection{RQ4. The Impact of Project Age}

\vspace{0.1cm}  \noindent {\bf Methodology:}
As the projects in the dataset are created in two different years, i.e., 2017 and 2018, we split the records into two groups based on the year in which a project is created. 
Then, we investigate the difference between developer productivity when working from home and working onsite for projects of different ages. 

\vspace{0.1cm}  \noindent {\bf Results:}
Table~\ref{tbl:age} presents the cases in which values when working from home are significantly different from those when working on site and the effect sizes are not negligible for projects created in 2017 and 2018.
As shown in this table, for projects created in 2017, there are more positive cases than negative cases. For example, the \emph{mean} of \emph{commit\_count}, \emph{release\_count}, and \emph{review\_count} are belong to positive cases while only the \emph{mean} of \emph{compile\_count} is belong to negative cases. 
On the other hand, all the cases of projects created in 2018 belong to negative cases. 
According to the feedback from Baidu, comparing to projects created in 2017, projects created in 2018 are less mature and might have more tasks and schedules. 
It is not easy for a project to complete some kinds of tasks or schedules when working at home, for example, recruiting a new developer usually takes more time when working at home, but a newer project usually needs more new developers than these older projects. 


\begin{tcolorbox}
    Working from home has a positive effect on developer productivity for projects created in 2017 but a negative impact on developer productivity for projects created in 2018.
\end{tcolorbox}

\begin{table}[]\vspace{-0.3cm}
    \centering
    \caption{The cases in which values when working from home is significantly different for {\tt \textbf{small}} and {\tt \textbf{large}} projects}\label{tbl:size}
    \begin{tabular}{@{}llcc@{}}
    \toprule
    Size                   & Metric          & Positive               & Negative    \\ \midrule
    \multirow{7}{*}{Small} & build\_count    & max                    & median, min \\
                           & commit\_count   & mean, sum, max         & ---         \\
                           & compile\_count  & ---                    & mean        \\
                           & lines\_deleted  & median                 & ---         \\
                           & lines\_inserted & median                 & ---         \\
                           & release\_count  & ---                    & ---         \\
                           & review\_count   & mean, median, sum, max & ---         \\ \midrule
    \multirow{7}{*}{Large} & build\_count    & sum                    & median, min \\
                           & commit\_count   & ---                    & ---         \\
                           & compile\_count  & ---                    & mean, sum   \\
                           & lines\_deleted  & ---                    & sum         \\
                           & lines\_inserted & ---                    & sum         \\
                           & release\_count  & ---                    & mean, max   \\
                           & review\_count   & ---                    & ---         \\ \bottomrule
    \end{tabular}
\end{table}

\begin{table*}[]\vspace{-0.3cm}
    \centering
    \caption{Developers whose productivity when working from home is significantly different from those when working onsite in terms of different features}\label{tbl:person}
    \begin{threeparttable}
    \resizebox{\textwidth}{!}{%
    \begin{tabular}{@{}llccccccc@{}}
    \toprule
    Project & Developer & build\_count    & commit\_count   & compile\_count      & lines\_deleted  & lines\_inserted & release\_count     & review\_count   \\ \midrule
    P1 & D1  & 0.567 (large)$^{***}$  & 0.634 (large)$^{***}$     & \xmark          & 0.324 (small)$^{**}$   & 0.295 (small)$^{*}$    & \xmark           & 0.452 (medium)$^{***}$ \\
P1 & D2  & \xmark            & -0.316 (small)$^{*}$      & -0.176 (small)$^{*}$ & \xmark            & \xmark            & \xmark           & \xmark            \\
P1 & D3  & \xmark            & \xmark               & \xmark          & \xmark            & \xmark            & \xmark           & 0.445 (medium)$^{***}$ \\
P1 & D4  & -0.395 (medium)$^{**}$ & -0.358 (medium)$^{**}$    & \xmark          & \xmark            & 0.272 (small)$^{*}$    & \xmark           & \xmark            \\
P1 & D5  & \xmark            & -0.083 (negligible)$^{*}$ & \xmark          & \xmark            & \xmark            & \xmark           & -0.335 (medium)$^{**}$ \\
P1 & D8  & \xmark            & \xmark               & \xmark          & -0.196 (small)$^{*}$   & -0.197 (small)$^{*}$   & 0.148 (small)$^{***}$ & \xmark            \\
P2 & D1  & \xmark            & -0.667 (large)$^{*}$      & \xmark          & \xmark            & -0.667 (large)$^{*}$   & \xmark           & 0.600 (large)$^{*}$    \\
P5 & D5  & -0.427 (medium)$^{**}$ & -0.296 (small)$^{*}$      & \xmark          & -0.290 (small)$^{*}$   & \xmark            & \xmark           & \xmark            \\
P5 & D6  & \xmark            & \xmark               & \xmark          & -0.293 (small)$^{*}$   & \xmark            & \xmark           & \xmark            \\
P5 & D7  & 0.585 (large)$^{**}$   & 0.530 (large)$^{**}$      & \xmark          & 0.367 (medium)$^{*}$   & 0.537 (large)$^{**}$   & \xmark           & \xmark            \\
P5 & D11 & \xmark            & -0.368 (medium)$^{*}$     & \xmark          & -0.405 (medium)$^{**}$ & -0.402 (medium)$^{**}$ & \xmark           & -0.299 (small)$^{*}$   \\
P5 & D13 & \xmark            & \xmark               & \xmark          & \xmark            & \xmark            & \xmark           & 0.288 (small)$^{**}$   \\
P5 & D14 & \xmark            & \xmark               & \xmark          & \xmark            & \xmark            & \xmark           & 0.214 (small)$^{*}$    \\
P5 & D17 & 1.000 (large)$^{**}$   & 0.636 (large)$^{*}$       & \xmark          & 0.591 (large)$^{*}$    & -0.773 (large)$^{**}$  & \xmark           & \xmark            \\
P5 & D20 & \xmark            & \xmark               & -0.167 (small)$^{*}$ & \xmark            & -0.451 (medium)$^{**}$ & \xmark           & \xmark            \\
P5 & D23 & \xmark            & \xmark               & \xmark          & 0.280 (small)$^{*}$    & \xmark            & \xmark           & 0.372 (medium)$^{**}$  \\
P5 & D24 & 0.351 (medium)$^{*}$   & 0.385 (medium)$^{**}$     & \xmark          & \xmark            & \xmark            & \xmark           & -0.368 (medium)$^{**}$ \\
P5 & D29 & \xmark            & \xmark               & \xmark          & \xmark            & \xmark            & -0.688 (large)$^{*}$  & \xmark            \\
P6 & D2  & \xmark            & \xmark               & \xmark          & \xmark            & \xmark            & \xmark           & -0.342 (medium)$^{*}$  \\
P7 & D2  & \xmark            & -0.312 (small)$^{*}$      & \xmark          & -0.413 (medium)$^{**}$ & -0.264 (small)$^{*}$   & \xmark           & \xmark            \\
P8 & D6  & \xmark            & \xmark               & \xmark          & \xmark            & \xmark            & \xmark           & 0.833 (large)$^{*}$     \\ \bottomrule
    \end{tabular}
    }
    \begin{tablenotes}\footnotesize
        \item[] * denotes p-value$<0.05$; ** denotes p-value$<0.01$; *** denotes p-value$<0.001$;.
    \end{tablenotes}
    \end{threeparttable}
    \vspace{-0.4cm}
\end{table*}

\begin{table}[]
    \centering
    \caption{The number of developers whose productivity when working from home is significantly larger and smaller than that when working onsite in terms of each metric.}\label{tbl:numberdeveloper}
    \begin{tabular}{@{}lll@{}}
    \toprule
    Metric         & \#Positive & \#Negative \\ \midrule
    build\_count   & 4 & 2 \\
    commit\_count  & 3 & 7 \\
    compile\_count & 0 & 2 \\
    line\_deleted  & 3 & 6 \\
    line\_inserted & 3 & 6 \\
    release\_count & 1 & 1 \\
    review\_count  & 7 & 4 \\ \bottomrule
    \end{tabular}
    \vspace{-0.4cm}
\end{table}

\subsection{RQ5. The Impact of Project Size}\label{sec:rq5}

\vspace{0.1cm}  \noindent {\bf Methodology:}
As shown in Table~\ref{tbl:projects}, these eight projects have different numbers of developers. According to the feedback from Baidu, we regard the project P4 and P5 whose number of developers are larger than 20 as {\tt large} projects, and the other 6 projects as {\tt small} projects. Then, we split the records into two groups based on project size. 
Then, we investigate the difference between developers' productivity when working from home and working onsite for projects with different sizes. 

\vspace{0.1cm}  \noindent {\bf Results:}
Table~\ref{tbl:size} presents the cases in which values when working from home are significantly different from those when working onsite and the effect sizes are not negligible for {\tt small} and {\tt large} projects.
We find that there are more positive cases than negative cases for {\tt small} projects and most of the cases in {\tt large} projects are belong to negative cases except for the \emph{sum} of \emph{build\_count}. 
The reason might be that it is more difficult for a large project to adjust their structure and schedule after working from home and it is more difficult for large projects to collaborate and communicate with others when working from home.


\begin{tcolorbox}
    Working from home has a larger impacts on developer productivity for {\tt large} projects than {\tt small} projects. 
\end{tcolorbox}

\subsection{RQ6. The Productivity of Individual Developers When Working From Home}\label{sec:rq6}

\vspace{0.1cm}  \noindent {\bf Methodology:}
For each developer in our dataset, we have two kinds of records, i.e., those when working from home and when working onsite. Then, we investigate whether their productivity when working from home is significantly different from that when working onsite in terms of each metric. We also compute Cliff's delta~\cite{cliff2014ordinal} to quantify the amount of difference. 
For the records when working on site, we only use {\tt DATA\_ONSITE} according to the findings in RQ1.

\vspace{0.1cm}  \noindent {\bf Results:}
Table~\ref{tbl:person} presents the individual developers if there exists at least one case in which the values of an aggregation of a metric when working from home are significantly different from those when working onsite. The second column in this table is the index of a developer in their project. 
Out of 139 developers in the whole dataset, the productivity of 21 developers when working from home is significantly different from when working onsite. 
On the other hand, the productivity of the majority of developers (84.8\%) in terms of all the metrics is not significantly different from when working onsite.


Table~\ref{tbl:numberdeveloper} presents the number of developers for who working from home has a positive or negative impact on their productivity in terms of a metric. 
As shown in this table, only in terms of \emph{compile\_count}, there is no developer whose productivity when working from home is significantly larger that that when working onsite. 
We also notice that the productivity of several developers when working from home is significantly larger than when working onsite in terms of all metrics, e.g., D1 of the project P1 and D7 of the project P5. 
For these developers, the company should encourage allow them to work from home for more time since remote work can improve their productivity. 
On the contrary, some developers are less productive when working from home, e.g., D2 of the project P1 and D5 of the project P5. For these developers, remote work is not encouraged since their productivity decreases when working from home.



\begin{tcolorbox}
    The productivity of the majority of developers when working from home is similar to that when working onsite. For a small portion of developers, working from home has different impacts on their productivity.
\end{tcolorbox}

\section{Discussion}\label{sec:diss}
In this section, we first present the feedback from Baidu, then provide implications of our findings. At the end of this section, we discuss some threats to validity.

\subsection{Feedback from Baidu}\label{sec:feedbacks}
Based on our findings, we perform a simple survey to get some feedback from developers in the studied projects. 
n the survey, we first collect some demographic information such as the developer role and the main programming language.
Then, we ask responders whether they agree that working from home has an impact on productivity in a  5-point Likert scale (\emph{strongly  disagree,disagree,neutral,agree,strongly  disagree}). 
Finally, we have a open question to ask them the factors that might affect developer productivity when working from home.
Many of them agreed that working from home can have both positive and negative impacts on developer productivity. Some also agree that there is no difference in productivity when working from home. 
The followings are some of the feedback we collected:

\vspace{0.1cm}  \noindent \textbf{Working from home improves developer productivity.}
{\fontsize{9}{12} 
\begin{itemize}
    \item[\leftthumbsup] \textit{It's \textbf{the first time} for some developers to work from home, so they feel very excited and have a lot of energy to do their work.}
    \item[\leftthumbsup] \textit{Developers can \textbf{focus on their own work} and not be disturbed by colleague. } 
    \item[\leftthumbsup] \textit{After working from home, the company asked developers to write \textbf{daily reports} instead of weekly reports. Daily reports can help developers recall their daily work and push them to work harder in the second day if their tasks aren't completed.}
    \item[\leftthumbsup] \textit{WFH decreases \textbf{the cost of transportation} and saves a lot of time for developers.} 
    \item[\leftthumbsup] \textit{WFH might increase developers' working time because there is \textbf{no switch} between workplace and home and developers can work very early in the morning or very late in the evening.} 
    \item[\leftthumbsup] \textit{WFH gives developers \textbf{better work-life balance} so that developers can work in better condition.} 
\end{itemize}
}

\vspace{0.1cm}  \noindent \textbf{Working from home decreases developer productivity.}
{\fontsize{9}{12}
\begin{itemize}
    \item[\leftthumbsdown] \textit{There are \textbf{much other stuff} (e.g., looking after children or pets, cooking by themselves, etc.) to interrupt developers' work and take a lot of their time.} 
    \item[\leftthumbsdown] \textit{Some developers \textbf{without self-discipline} cannot focus on work when working from home. Unlikely onsite, they might be too relaxed at home since there is no colleague around them.}
    \item[\leftthumbsdown] \textit{Although \textbf{video conferencing tools or telephone} are now very convenient for communication, \textbf{the efficiency of collaboration} still decreases due to working from home.}
\end{itemize}
}

\vspace{0.1cm}  \noindent \textbf{There is no difference in developer productivity when working from home.}
{\fontsize{9}{12}
\begin{itemize}
    \item[\HandRight] \textit{There are no barriers for many developers to complete their daily tasks (e.g., writing code, build projects, code review, etc.) when working from home.}
    \item[\HandRight]  \textit{There is no difference of project schedule between working from home and working onsite since developers can know the schedule using an online project schedule tool.}
    \item[\HandRight]  \textit{Current video conferencing tools are very powerful, for example, they usually support screen sharing. So, there is a very slight difference between meeting at a meeting room of the company and online.}
\end{itemize}
}

\subsection{Implications}
\vspace{0.1cm}  \noindent \textbf{Working from home has different impacts on overall developer productivity.} 
Many previous studies have shown that working from home has a positive effect on the productivity of workers~\cite{neufeld2004predicting, neufeld2005individual, baker2007satisfaction, laihonen2012measuring, coenen2014workplace, campbell2015flexible, kazekami2020mechanisms}. 
Some other studies also showed that working from home might have a negative impact on employee productivity. For example,  Kazekami found that the long working time of WFH would decrease teleworker productivity~\cite{kazekami2020mechanisms}.  
In our study, we use a quantitative analysis method to show that working from home has different impacts on developer productivity. From the feedback of Baidu, the difference in developer productivity might be caused by many reasons. 
We also find that the productivity of the majority of developers is not significantly different from when working onsite (RQ6), so we think WFH can be considered as a choice of work arrangement for employees because WFH offers many other benefits except productivity, such as saving costs for the company and the flexibility of working time for developers. 


\vspace{0.1cm}  \noindent \textbf{A project needs to prepare for working from home according to its own characteristics.} 
In this study, we find that developer productivity might be associated with the characteristics of a project including programming language and project type/age/size. 
For example, working from home might have a positive effect on developer productivity for {\tt small} project but does not affect developer productivity for {\tt large} projects (see RQ2 in Section~\ref{sec:rq5}).  
Thus, we believe that adopting the working from home policy for a project should be based on its own characteristics, e.g., programming languages, and project size. 
When starting the working from home policy, a project needs to prepare some resources to reduce the risks of decreasing developer productivity. For example, a large project should consider the communication cost of WFH and prepare the relevant tools to facilitate the communication of the team.

\vspace{0.1cm}  \noindent \textbf{Using different strategies of working from home for individual developers.}
We find that the productivity of the majority of developers in this study does not change when working from home. Still there exist some developers whose productivity when working from home is different from these when working onsite.
Thus, some approaches based on some development metrics can be used to identify whether the productivity of a developer increases or decreases when working from home. Once the productivity of a developer decreases, the project manager needs to identify the reasons behind it. If the developer is not suitable to WFH, they should be asked to go to the company for work. 
On the other hand, if the productivity of a developer increases, the project team should allow them to continue working from home. 

For researchers, to improve individual developers' productivity when working from home, more empirical studies are required to investigate more factors that affect their productivity, e.g., personality, moods, and their working environment at home. Additionally, some machine learning models based on developers' daily activities can be built to predict whether their productivity will change when working from home.

\subsection{Threats to Validity}

\vspace{0.1cm}  \noindent \textbf{Threats to internal validity.} First, there might exist errors in our code and experiment setting. We have written a python script to process and analyze the dataset provided by Baidu. We double-check our code, however, there may exist some errors that we do not notice.
The second internal validity is that we use some quantitative metrics of software development (e.g., the number of builds and commits) in the dataset to measure developer productivity. 
These metrics such as lines of code have been used to measure developer productivity~\cite{walston1977method, devanbu1996analytical, nguyen2011analysis}.
Hence, we think these metrics can potentially indicate the productivity of developers and we also use multiple aggregate values (such as \emph{mean} and \emph{median}) of these metrics by day to measure the productivity. 
Third, there might be many other factors (e.g., the workload in different times) that affect developer productivity. It is difficult to exclude all other factors in the study. To minimize this threat, we use the records of developers' activities of WFH and ONSITE in the same year (i.e., 2020) for comparison.
Finally, the metrics in the dataset are standardized due to the security policy and privacy of Baidu. But we focus on the difference between the productivity of developers when working from home and working onsite. Thus, we believe that the comparison results using standardized values does not affect the findings in the study.

\vspace{0.1cm}  \noindent \textbf{Threats to external validity} relate to the generalizability of our findings. In this study, the dataset we used is from Baidu. The number of projects and developers are limited. Thus, it is unclear whether the same results still hold for other developers from other companies. 
However, since Baidu is one of the largest IT companies in China, we believe that our findings in this study have typical and common characteristics to some extent.
Additionally, we analyze 138 working days of software development activities from 139 developers. These developers are from eight different projects with different characteristics such as programming languages and project types. 
Another threat to external validity relate to the generalizability of the metrics used to measure the productivity of developers. These metrics used in this study are very general and often used in software development~\cite{bao2019large, Yan2019Characterizing}. 
In the future, to reduce these threats, we plan to investigate more developers from different companies and consider more metrics.

\section{Related Work}\label{sec:related}
In this section, we discuss related work in fields of working from home and developer productivity.

\vspace{-0.3cm}
\subsection{Working from Home}
There are many studies in the literature that investigate the benefits and drawbacks of WFH~\cite{nilles1994making, bailey1999advantages, shin2000telework, perez2002benefits, felstead2017assessing, aguilera2016home}. 
According to the literature, the main benefits of WFH for companies include saving costs of buildings and increasing productivity and job satisfaction of employees. For employees, WFH gives them more flexible working time and provides better work-life balance. 
WFH can also offer benefits to some special kinds of persons, such as the disabled~\cite{spark2017accessibility} and transgender developers~\cite{ford2019remote}.
The main disadvantages of WFH are the access to technology and the integration of telework with the company's strategy and organisational structure, as well as the teleworkers motivation and control~\cite{perez2002benefits}. Felstead and Henseke also reported that telework makes employees it difficulty to insulate the world of work from other aspects of life when both worlds collide and overlap~\cite{felstead2017assessing}.

Several studies have investigated the impact of WFH on productivity~\cite{neufeld2004predicting, neufeld2005individual, baker2007satisfaction, laihonen2012measuring, coenen2014workplace, campbell2015flexible, kazekami2020mechanisms}.
Among these studies, many of them reported that WFH has a positive impact on the productivity of teleworkers.  
For example, Coenen and Kok found that telework has a positive effect on the performance of new product development through enabling knowledge sharing, cross-functional cooperation and inter-organizational involvement~\cite{campbell2015flexible}. 
On the contrary, WFH might have a negative impact on productivity. For instance, Kazekami found that appropriate telework hours increase labor productivity but when telework hours are too long, telework decreases labor productivity~\cite{kazekami2020mechanisms}. WFH might decrease the efficiency of developer communication, which plays an important role in software development~\cite{wolf2009predicting}.

However, most of the previous studies for WFH are based on a qualitative analysis using survey or interviews. The participants in these studies are general workers not only developers, and they do not have the experience of working from home for a long continuing time. 
In this study, we perform a quantitative analysis based on a lot of activity data of developers when working from home during the time of COVID-19 pandemic. We focus on the impact of WFH on developer productivity and the potential factors affecting developer productivity. 


\subsection{Developer Productivity}
A lot of studies use developers' daily activities to investigate their productivity. 
For example, Perry et al. found that many developers spend a lot of time on the communication with colleagues~\cite{perry1994people}.  
Additionally, many studies reported that developers work is fragmented and frequently interrupted, which has an important impact on their productivity~\cite{bailey2001effects, chong2006interruptions, czerwinski2004diary, horvitz2001notification, parnin2010evaluating, van1998interrupts}. For example, Sanchez et al. found that work fragmentation is correlated to lower observed productivity and longer activity switches seem to strengthen the effect~\cite{sanchez2015empirical}. 

Developer productivity is often measured by software artifacts produced by developers in a certain time, e.g., submitted lines of code (LOC)~\cite{devanbu1996analytical, nguyen2011analysis}, function points~\cite{albrecht1979measuring}, completed tasks~\cite{minelli2015know}, and time to implement a requirement~\cite{cataldo2008socio}. Meyer et al. proposed
a list of metrics to measure developer productivity and several ways to improve a developer's productivity through tool support~\cite{meyer2014software}.
Some studies also investigate the factors affecting developer productivity, e.g., characteristics of workplace (e.g., privacy, noise)~\cite{demarco1985programmer}, programming languages and development tools~\cite{boehm1987improving}, project switching~\cite{vasilescu2016sky}, and developers' mood~\cite{khan2011moods}. Additionally, personal factors might have an impact on productivity, for example, some developers feel more productive when communicating with others but some don't like to be interrupted when working~\cite{meyer2017characterizing}. 

Due to the outbreak of COVID-19, some researchers also start to investigate the effect of the pandemic on developers' productivity. 
For example, Ralph et al. conducted a survey and found that the pandemic has had a negative effect on developers’ productivity~\cite{ralph2020pandemic}. 

In this study, we focus on the difference of developer productivity between working from home and working onsite. We measure developer productivity by several metrics based on developers' daily activities, which have been used in previous studies. We also investigate the factors affecting developer productivity when working from home, such as programming language, project type, project size, etc. 


\section{Conclusion}\label{sec:conclusion}
In this paper,  we investigate the productivity of developers when working from home for a long time due to the COVID-19 pandemic. 
We use a quantitative analysis based on a dataset of developers' daily activities from Baidu.
To compare developer productivity when working from home with that when working on site, we use several metrics of software development in the dataset, such as the number of builds, commits, and inserted/deleted lines. 
We find that working from home has different impacts on developer productivity in terms of different metrics.
Also, we investigate some factors affecting developer productivity when working from home including programming language and project type/age/size. 
Additionally, we find that a small number of individual developers have different productivity when working from home.
In the future, we plan to extend our study by using more data from more developers and companies. We also want to build machine learning models to predict developer productivity based on developers' daily activities.

\Acknowledgements{This research was partially supported by the National Key Research and Development Program of China(2018YFB1003904), NSFC Program (No. U20A20173 and No. 61902344)), the Natural Science Foundation of Zhejiang Province (no. LY21F020011).}

\balance
\bibliographystyle{scis}
\bibliography{reference}

\end{document}